\documentclass[showkeys,showpacs,superscriptaddress]{revtex4}
\usepackage{amsmath}
\usepackage{amssymb}
\usepackage{graphicx}
\usepackage{color,ulem}

\begin{document}

\title{Dirac stars supported by nonlinear spinor fields
}
\author{Vladimir Dzhunushaliev}
\email{v.dzhunushaliev@gmail.com}
\affiliation{
	Institute of Experimental and Theoretical Physics,  Al-Farabi Kazakh National University, Almaty 050040, Kazakhstan
}
\affiliation{
National Nanotechnology Laboratory of Open Type,  Al-Farabi Kazakh National University, Almaty 050040, Kazakhstan
}
\affiliation{
	Department of Theoretical and Nuclear Physics,  Al-Farabi Kazakh National University, Almaty 050040, Kazakhstan
}
\affiliation{
	Institute of Physicotechnical Problems and Material Science of the NAS of the Kyrgyz Republic, 265 a, Chui Street, Bishkek 720071,  Kyrgyzstan
}
\affiliation{
	Institut f\"ur Physik, Universit\"at Oldenburg, Postfach 2503
	D-26111 Oldenburg, Germany
}

\author{Vladimir Folomeev}
\email{vfolomeev@mail.ru}
\affiliation{
	Institute of Experimental and Theoretical Physics,  Al-Farabi Kazakh National University, Almaty 050040, Kazakhstan
}
\affiliation{
National Nanotechnology Laboratory of Open Type,  Al-Farabi Kazakh National University, Almaty 050040, Kazakhstan
}
\affiliation{
	Institute of Physicotechnical Problems and Material Science of the NAS of the Kyrgyz Republic, 265 a, Chui Street, Bishkek 720071,  Kyrgyzstan
}
\affiliation{
	Institut f\"ur Physik, Universit\"at Oldenburg, Postfach 2503
	D-26111 Oldenburg, Germany
}


\begin{abstract}
We study configurations consisting of a gravitating spinor field $\psi$ with
a nonlinearity of the type $\lambda\left(\bar\psi\psi\right)^2$.
To ensure spherical symmetry of the configurations, we use two spin-$\frac{1}{2}$ fields forming a spin singlet.
For such systems, we find regular stationary asymptotically flat solutions describing
compact objects. For negative values of the coupling constant $\lambda$, it is shown that,
by choosing physically reasonable values of this constant,
it is possible to obtain configurations with masses comparable to the Chandrasekhar mass.
It enables us to speak of an astrophysical interpretation of the obtained systems, regarding them as Dirac stars.
\end{abstract}



\maketitle

\section{Introduction}

In recent decades, various fundamental fields have achieved widespread use in a variety of cosmological  and astrophysical applications.
In particular, this applies both to modeling the present accelerated expansion of the Universe~\cite{AmenTsu2010}
and to describing its early inflationary stage~\cite{Linde:2007fr}. For this purpose, various scalar (boson) fields with spin 0 are most frequently employed.
Such fields are also widely used in modeling compact astrophysical  strongly gravitating objects~-- boson stars~\cite{Schunck:2003kk}.

However, there may exist gravitating objects consisting of fields with nonzero spin. They may be systems
supported by fields with integer spin: Yang-Mills configurations~\cite{Bartnik:1988am} (consisting of massless vector fields) or
Proca stars~\cite{Brito:2015pxa} (consisting of massive vector fields). In the case of spin-$\frac{1}{2}$ fields,
 the literature in the field offers both gravitating configurations with noninteracting spinor fields~\cite{Finster:1998ws,Herdeiro:2017fhv} and
 objects supported by nonlinear fields. In particular, nonlinear spinor fields have been used in obtaining cylindrically symmetric solutions
 in Ref~\cite{Bronnikov:2004uu} (stringlike configurations) and in Ref.~\cite{Bronnikov:2009na} (wormhole solutions)
 and also in a cosmological context in Refs.~\cite{Saha:2014hea,Saha:2015sna,Saha:2015rna,Saha:2016ebv,Saha:2016cbu},
 where the role of spinor fields in the evolution of
  anisotropic universes described by the Bianchi type  I, III, V, VI, and VI$_0$ models or of an isotropic Friedmann-Robertson-Walker universe is studied.
 In turn, for spherically symmetric systems,  localized regular solutions have been found in Refs.~\cite{Krechet:2014nda,Adanhounme:2012cm}.
 The papers~\cite{Ribas:2010zj,Ribas:2016ulz}
 study models of the universe filled with tachyon and fermion fields interacting through the Yukawa scalar field.
 In Ref.~\cite{Dzhunushaliev:2018bdi}, a topologically nontrivial solution with a spinor field within the Einstein-Dirac theory has been obtained.

Configurations consisting of spinor fields are prevented from collapsing under their own gravitational fields due to the Heisenberg uncertainty principle.
The distinctive feature of such systems is that, since the spin of a fermion has an intrinsic orientation in space, a system consisting of a single
spinor particle cannot possess spherical symmetry. In order to ensure the latter, one can take two fermions having opposite spin, i.e., consider two spinor fields.
For each of such spinors, the energy-momentum tensors will not be spherically symmetric (due to the existence of nondiagonal components), but
their sum will give a tensor compatible with spherical symmetry of the spacetime
 (see below in Sec.~\ref{prob_statem}).

In the case of configurations supported by noninteracting spinor fields, their total mass
$M\sim M_p^2/\mu$, where $\mu$ is the spinor field mass, and it is generally much smaller than the Chandrasekhar mass,
$M_{\text{Ch}}\sim M_p^3/\mu^2$.  In the present paper, we consider the case of nonlinear spinor fields and show that in this case
there is a possibility of increasing the total mass considerably.

We emphasize here that in the present paper we will consider a system supported by a {\it classical} spinor field. Following Ref.~\cite{ArmendarizPicon:2003qk},
by the latter, we mean a set of four complex-valued spacetime functions that transform according to
the spinor representation of the Lorentz group. But it is obvious that realistic spin-$\frac{1}{2}$ particles must be described
by {\it quantum} spinor fields. It is usually believed that there exists no classical limit for quantum spinor fields.
However, classical spinors can be regarded as arising from some effective description of more complex quantum systems (for possible justifications of the existence of classical spinors,
see Ref.~\cite{ArmendarizPicon:2003qk}).

The paper is organized as follows. In Sec.~\ref{prob_statem}, we present the general-relativistic equations for the systems under consideration.
These equations are solved numerically in Sec.~\ref{num_sol} for different values of
the coupling constant $\lambda$, and the possibility of obtaining configurations with astrophysical masses of the order of the Chandrasekhar mass
is demonstrated. Finally, in Sec.~\ref{concl}, we summarize and discuss the obtained results.

\section{Formulating the problem and general equations}
\label{prob_statem}

We consider compact gravitating configurations consisting of a spinor field and modeled within the framework of Einstein's general relativity.
The corresponding action for such a system can be represented in the form
[the metric signature is $(+,-,-,-)$]
\begin{equation}
\label{action_gen}
	S=-\frac{c^3}{16\pi G}\int d^4 x \sqrt{-g} R +S_{\text{sp}},
\end{equation}
where $G$ is the Newtonian gravitational constant,
$R$ is  the scalar curvature, and $S_{\text{sp}}$ denotes the action
of the spinor field.  This action is obtained from the Lagrangian for the spinor field  $\psi$ with the mass $\mu$,
\begin{equation}
	L_{\text{sp}} =	\frac{i \hbar c}{2} \left(
			\bar \psi \gamma^\mu \psi_{; \mu} -
			\bar \psi_{; \mu} \gamma^\mu \psi
		\right) - \mu c^2 \bar \psi \psi - F(S),
\label{lagr_sp}
\end{equation}
which contains the covariant derivatives
$
\psi_{; \mu} = \left[\partial_{ \mu} +1/8\, \omega_{a b \mu}\left( \gamma^a  \gamma^b- \gamma^b  \gamma^a\right)\right]\psi
$,
and $\gamma^a$ are the Dirac matrices in the standard representation in flat space
 [see, e.g.,  Ref.~\cite{Lawrie2002}, Eq.~(7.27)]. In turn, the Dirac matrices in curved space,
$\gamma^\mu = e_a^{\phantom{a} \mu} \gamma^a$, are obtained using the tetrad
 $ e_a^{\phantom{a} \mu}$, and $\omega_{a b \mu}$ is the spin connection
[for its definition, see Ref.~\cite{Lawrie2002}, Eq.~(7.135)].
Finally, this Lagrangian contains an arbitrary nonlinear term $F(S)$, where the invariant $S$ can depend on
$
	\left( \bar\psi \psi \right),
	\left( \bar\psi \gamma^\mu \psi \right)
	\left( \bar\psi \gamma_\mu \psi \right)$ or
	$\left( \bar\psi \gamma^5 \gamma^\mu \psi \right)
	\left( \bar\psi \gamma^5 \gamma_\mu \psi \right)$.
Here, we will study the case of the simplest nonlinearity
$F(S) \propto \left( \bar\psi \psi \right)^2$.

Varying the action \eqref{action_gen} with respect to the metric and the spinor field, we derive the Einstein equations and the Dirac equation in curved spacetime:
\begin{eqnarray}
	R_{\mu}^\nu - \frac{1}{2} \delta_{\mu }^\nu R &=&
	\frac{8\pi G}{c^4} T_{\mu }^\nu,
\label{feqs-10} \\
	i \hbar \gamma^\mu \psi_{;\mu} - \mu c \psi - \frac{1}{c}\frac{\partial F}{\partial\bar\psi}&=& 0,
\label{feqs-20}\\
	i \hbar \bar\psi_{;\mu} \gamma^\mu + \mu c \bar\psi +
	\frac{1}{c}\frac{\partial F}{\partial\psi}&=& 0.
\label{feqs-21}
\end{eqnarray}
The right-hand side of Eq.~\eqref{feqs-10} contains the spinor field energy-momentum tensor
 $T_{\mu}^\nu$, which can be represented (already in a symmetric form) as
\begin{equation}
	T_{\mu}^\nu =\frac{i\hbar c }{4}g^{\nu\rho}\left[\bar\psi \gamma_{\mu} \psi_{;\rho}+\bar\psi\gamma_\rho\psi_{;\mu}
-\bar\psi_{;\mu}\gamma_{\rho }\psi-\bar\psi_{;\rho}\gamma_\mu\psi
\right]-\delta_\mu^\nu L_{\text{sp}}.
\label{EM}
\end{equation}
Next, taking into account the Dirac equations \eqref{feqs-20} and \eqref{feqs-21}, the Lagrangian \eqref{lagr_sp} takes the form
$$
	L_{\text{sp}} = - F(S) + \frac{1}{2} \left(
		\bar\psi\frac{\partial F}{\partial\bar\psi} +
		\frac{\partial F}{\partial\psi}\psi
	\right).
$$
For our purpose,  we choose the nonlinear term appearing in this Lagrangian in a simple power-law form,
\begin{equation}
	F(S) = - \frac{k}{k+1}\lambda\left(\bar\psi\psi\right)^{k+1},
\label{nonlin_term}
\end{equation}
where $k, \lambda$ are some free parameters. Below, we take $k=1$ to yield
\begin{equation}
	F(S) = - \frac{\lambda}{2} \left(\bar\psi\psi\right)^2.
\label{nonlin_term_2}
\end{equation}
[Regarding the physical meaning of the constant $\lambda$, see Eq.~\eqref{p_r_eff} below.]

Since in the present paper we consider only spherically symmetric configurations, it is convenient to choose the spacetime metric in the form
\begin{equation}
	ds^2 = N(r) \sigma^2(r) (dx^0)^2 - \frac{dr^2}{N(r)} - r^2 \left(
		d \theta^2 + \sin^2 \theta d \varphi^2
	\right),
\label{metric}
\end{equation}
where $N(r)=1-2 G m(r)/(c^2 r)$, and the function $m(r)$ corresponds to the current mass of the configuration
enclosed by a sphere with circumferential radius $r$; $x^0=c t$ is the time coordinate.

In order to describe the spinor field, it is necessary to choose the corresponding ansatz for $\psi$
compatible with the spherically symmetric line element \eqref{metric}.
This can be done as follows (see, e.g., Refs.~\cite{Li:1982gf,Li:1985gf,Herdeiro:2017fhv}),
\begin{equation}
	\psi^T =2\, e^{-i \frac{E t}{\hbar}} \begin{Bmatrix}
		\begin{pmatrix}
			0 \\ - g \\
		\end{pmatrix},
		\begin{pmatrix}
			g \\ 0 \\
		\end{pmatrix},
		\begin{pmatrix}
			i f \sin \theta e^{- i \varphi} \\ - i f \cos \theta \\
		\end{pmatrix},
		\begin{pmatrix}
			- i f \cos \theta \\ - i f \sin \theta e^{i \varphi} \\
		\end{pmatrix}
	\end{Bmatrix},
\label{spinor}
\end{equation}
where $E/\hbar$ is the spinor frequency,  $f(r)$ and $g(r)$ are two real functions.
 This ansatz ensures that the spacetime of the system under consideration remains static. 
 Here, each row describes a fermion with spin $1/2$, and these two fermions have opposite spins. 
 Also, each row corresponds to the spinor ansatz of Ref.~\cite{Finster:1998ws} and it is related to the spinor ansatz of Ref.~\cite{Herdeiro:2017fhv} by the expression
$
\psi_{[7]} = S \psi_i,
$
where $i = 1,2$ is the row number from \eqref{spinor}. The matrix $S$ is given by the expression~\cite{obukhov}
\begin{equation}
	S = \frac{1}{\sqrt 2} \begin{pmatrix}
		e^{i (\theta + \varphi)/2} & - i e^{i (\theta - \varphi)/2} &	0	&	0 	\\
		e^{- i (\theta - \varphi)/2} & i e^{- i (\theta + \varphi)/2} &	0	&	0 \\
		0 & 0 &	e^{i (\theta + \varphi)/2}	&	- i e^{i (\theta - \varphi)/2} 	\\
		0	&	0 &	e^{- i (\theta - \varphi)/2} & i e^{- i (\theta + \varphi)/2} \\
	\end{pmatrix} .
\label{matrix_S}
\end{equation}
Thus, the ansatz  \eqref{spinor} describes two Dirac fields,
and for each of them, the energy-momentum tensor is not spherically symmetric, but their sum yields a spherically symmetric energy-momentum tensor.

Now, substituting the ansatz \eqref{spinor} [by multiplying each row of \eqref{spinor} by the matrix \eqref{matrix_S}]
and the metric \eqref{metric} into the field equations \eqref{feqs-10} and \eqref{feqs-20}, we have
\begin{eqnarray}
	&&\bar f^\prime + \left[
		\frac{N^\prime}{4 N} + \frac{\sigma^\prime}{2\sigma}+\frac{1}{x}\left(1+\frac{1}{\sqrt{N}}\right)
	\right] \bar f + \left(
		\frac{1}{\sqrt{N}} - \frac{\bar E}{\sigma N} +
		8\bar \lambda\,\frac{\bar f^2 - \bar g^2}{\sqrt{N}}
	\right)\bar g= 0,
\label{fieldeqs-1_dmls}\\
	&&\bar g^\prime + \left[
			\frac{N^\prime}{4 N} + \frac{\sigma^\prime}{2\sigma} +
			\frac{1}{x}\left(1 - \frac{1}{\sqrt{N}}\right)
	\right]\bar g + \left(
		\frac{1}{\sqrt{N}} + \frac{\bar E}{\sigma N} +
		8\bar \lambda\,\frac{\bar f^2 - \bar g^2}{\sqrt{N}}
	\right)\bar f= 0,
\label{fieldeqs-2_dmls}\\
	 &&\bar m^\prime=8 x^2\left[
\bar E\,\frac{\bar f^2+\bar g^2}{\sigma\sqrt{N}}+4\bar \lambda\left(\bar f^2-\bar g^2\right)^2
\right],
\label{fieldeqs-3_dmls}\\
&&\frac{\sigma^\prime}{\sigma}	=\frac{8 x}{\sqrt{N}}\left[
\bar E\,\frac{\bar f^2+\bar g^2}{\sigma N}+ \bar g \bar f^\prime-\bar f \bar g^\prime
\right],
\label{fieldeqs-4_dmls}
\end{eqnarray}
where the prime denotes differentiation with respect to the radial coordinate.
Here,  Eqs.~\eqref{fieldeqs-3_dmls} and \eqref{fieldeqs-4_dmls} are the  ($^0_0$) and  $[(^0_0)~-~(^1_1)]$ components of the Einstein equations,
respectively. The above equations are written in terms of the following dimensionless variables,
\begin{equation}
\label{dmls_var}
	x = r/\lambda_c, \quad
	\bar E = \frac{E}{\mu c^2}, \quad
	\bar f, \bar g = \sqrt{4\pi}\lambda_c^{3/2}\frac{\mu}{M_p} f, g,\quad
	\bar m = \frac{\mu}{M_p^2} m, \quad
	\bar \lambda = \frac{1}{4\pi \lambda_c^3\mu c^2}
	\left(\frac{M_p}{\mu}\right)^2\lambda,
\end{equation}
 where $M_p$ is the Planck mass and $\lambda_c=\hbar/\mu c$ is the constant having
the dimensions of length (since we consider a classical theory, $\lambda_c$ need not be associated with the Compton length); the metric function
 $N=1-2\bar m/x$. Note here that, using the Dirac equations \eqref{fieldeqs-1_dmls} and \eqref{fieldeqs-2_dmls},
 one can eliminate the derivatives of $\bar f$ and $\bar g$ from the right-hand side of Eq.~\eqref{fieldeqs-4_dmls}.

For numerical integration of the above equations, we  take
the following boundary conditions in the vicinity of the center,
$$
	\bar g\approx \bar g_c + \frac{1}{2}\bar g_2 x^2, \quad
	\bar f\approx \bar f_1 x,
	\quad \sigma\approx \sigma_c+\frac{1}{2}\sigma_2 x^2, \quad
	\bar m\approx \frac{1}{6}\bar m_3 x^3,
$$
where the index c denotes central values of the corresponding variables.
The expansion coefficients  $\bar f_1,  \bar m_3, \sigma_2, \bar g_2$ can be found from the set of Eqs.~\eqref{fieldeqs-1_dmls}-\eqref{fieldeqs-4_dmls}. 
In turn, the expansion coefficients
$\sigma_c$ and $\bar g_c$, and also the parameter $\bar E$,
are arbitrary. Their values are chosen so as to obtain regular and asymptotically flat solutions with
the functions $N(x\to \infty),\sigma(x\to \infty) \to 1$.
In this case, the asymptotic value of the function
$\bar m$ will correspond to the Arnowitt-Deser-Misner (ADM) mass of the configurations under consideration.

\section{Numerical solutions}
\label{num_sol}

\begin{figure}[t]
	\begin{minipage}[t]{.49\linewidth}
		\begin{center}
			\includegraphics[width=1\linewidth]{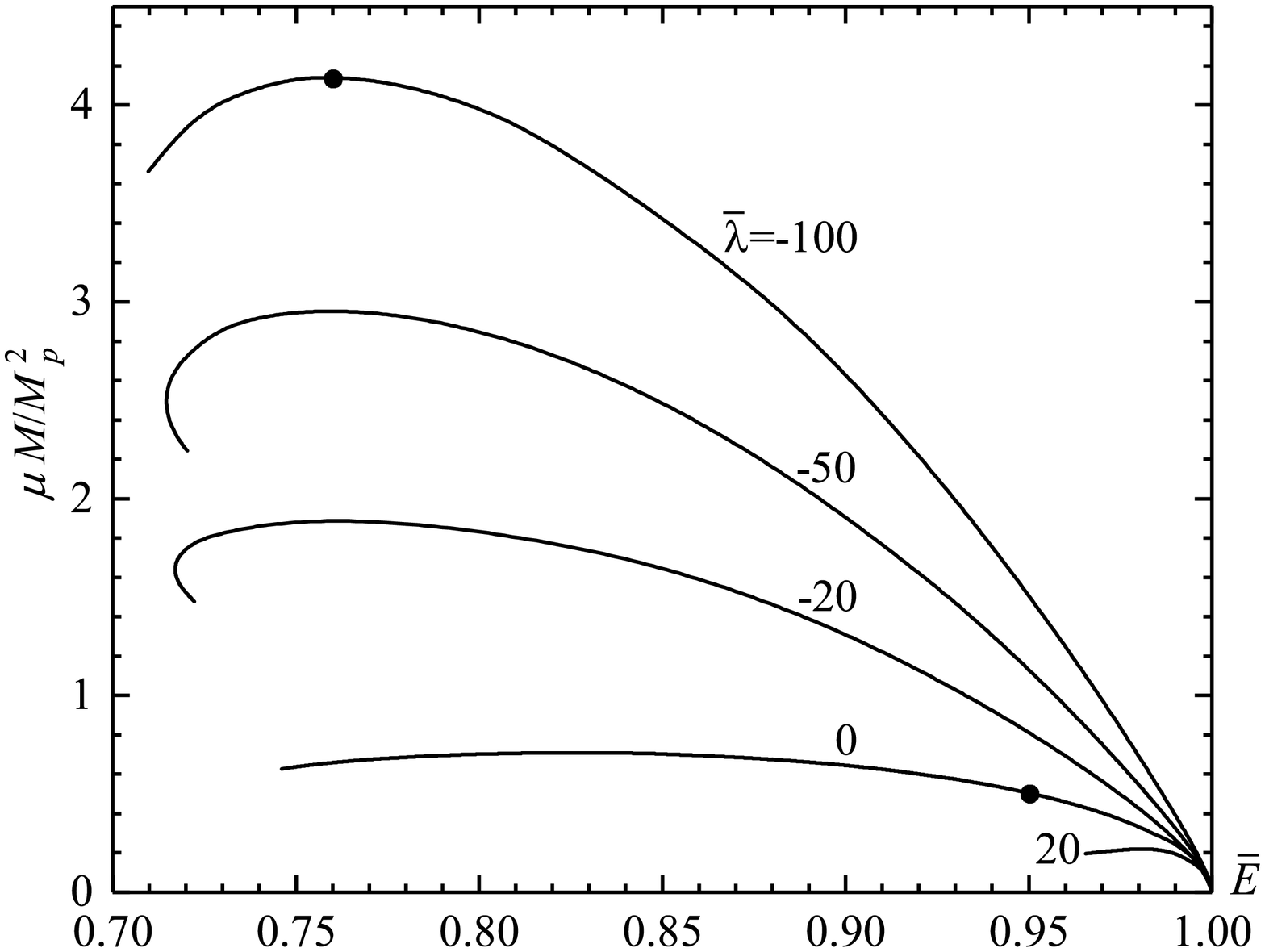}
		\end{center}
\vspace{-0.5cm}
		\caption{Dimensionless Dirac-star total mass $\bar M$ as a function of the parameter $\bar E$ for $\bar \lambda=-100, -50, -20, 0,$ and $20$.
The bold dots mark the positions of the configurations for which the graphs of Fig.~\ref{fig_field_distr} are plotted.
		}
		\label{fig_mass_E}
	\end{minipage}\hfill
\begin{minipage}[t]{.49\linewidth}
		\begin{center}
			\includegraphics[width=1\linewidth]{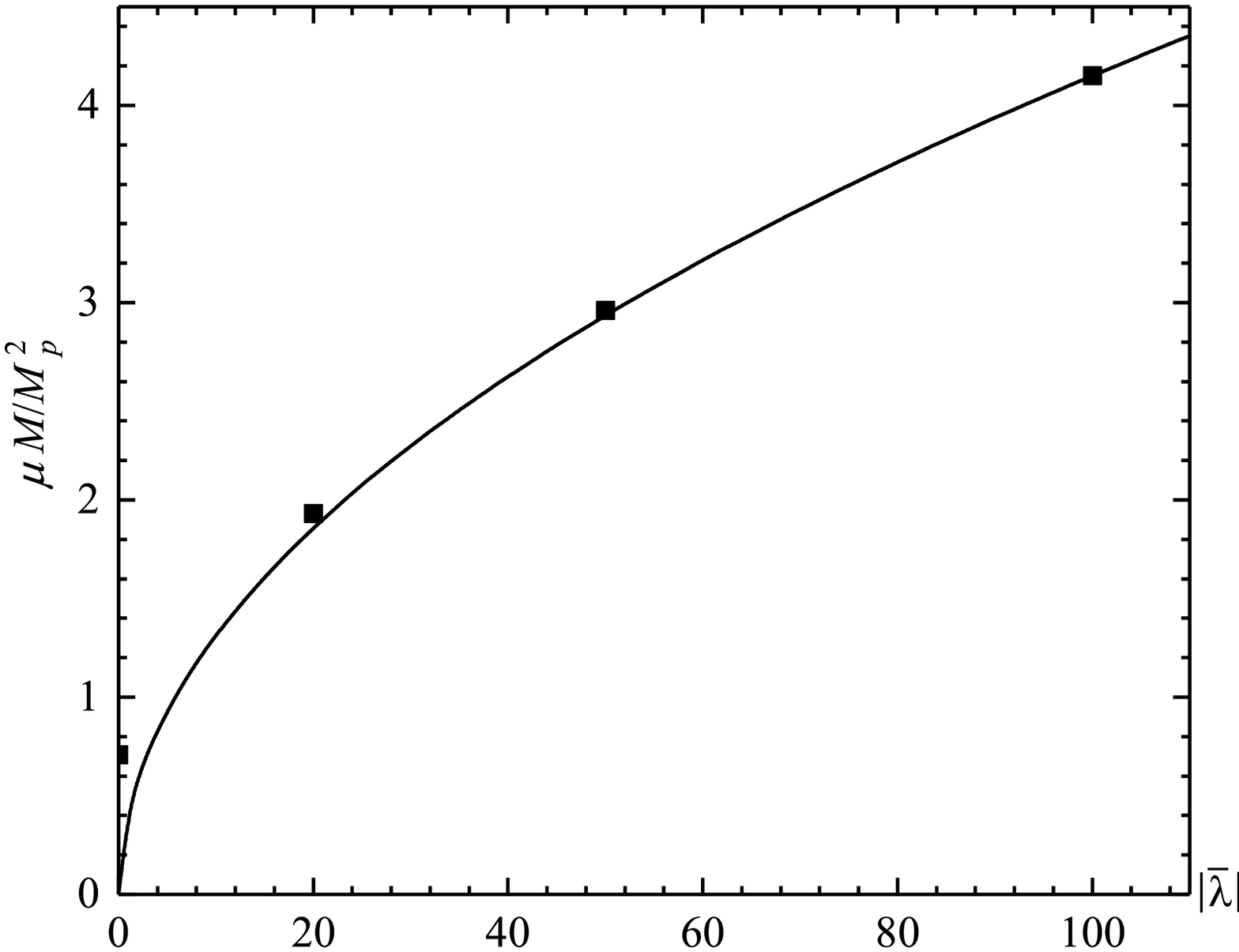}
		\end{center}
\vspace{-0.5cm}
		\caption{Maximum Dirac-star masses as a function of $|\bar \lambda|$. The solid curve corresponds to the asymptotic relation \eqref{M_max_approx}.
		}
		\label{fig_mass_lambda}
	\end{minipage}
\end{figure}

Integration of  Eqs.~\eqref{fieldeqs-1_dmls}-\eqref{fieldeqs-4_dmls} is performed  from the center of the configuration
(at $x\approx 0$), where a particular value of  $\bar g_c$
 corresponding to the central density of the spinor field is specified, to some boundary point where the functions
$\bar g, \bar f $ and their derivatives go to zero. Since with increasing distance the spinor fields decrease exponentially fast as
$
\bar g, \bar f \sim e^{-\sqrt{1-\bar E^2}\,x}
$,
this point can be approximately regarded as some effective radius $x_{\text{eff}}$ of the configurations under investigation
(by analogy with the case of boson stars~\cite{Schunck:2003kk}). Depending on the value of the central density of the spinor field,
$x_{\text{eff}}$ is of the order of several hundreds for $\bar g_c \approx 0$,
and it decreases down to $x_{\text{eff}}\sim 10$ for $\bar g_c \sim 1$; i.e., as the central density increases, the characteristic sizes of the configurations under consideration decrease.
In turn, the parameter $\bar E$, starting from the value
 $\bar E \approx 1$ for $\bar g_c \approx 0$, at first decreases as $\bar g_c$ increases
 and then can start growing again. This is illustrated in Fig.~\ref{fig_mass_E} where the dependencies of the Dirac-star total mass
 $\bar M$ on $\bar E$ are shown for different values of the coupling constant $\bar \lambda$.

In plotting the above dependencies, we have kept track of the sign of the binding energy (BE),
which is defined as the difference between the energy of $N_f$ free particles, ${\cal E}_f=N_f \mu c^2$, and the total energy of the system, ${\cal E}_t=M c^2$,
i.e., $\text{BE}={\cal E}_f-{\cal E}_t$. Here, the total particle number
 $N_f$ is equal to the Noether charge $Q$ of the system, which is defined via the timelike component of the 4-current $j^\alpha=\sqrt{-g}\bar \psi \gamma^\alpha \psi$
as
$
Q=\int j^t d^3 x,
$
where in our case $j^t = N^{-1/2}r^2 \sin{\theta} \left(\psi^\dag \psi\right)$.
In the dimensionless variables \eqref{dmls_var}, we then have
$$
N_f=Q=8\left(\frac{M_p}{\mu}\right)^2\int_0^\infty \frac{\bar f^2+\bar g^2}{\sqrt{N}}x^2 dx.
$$
A necessary, albeit not sufficient, condition for energy stability
is the positiveness of the binding energy.
Therefore, since configurations with a negative BE are certainly unstable, the graphs in Fig.~\ref{fig_mass_E}
are plotted only up to $\bar E$ for which the BE becomes equal to 0
(except the case of $\bar \lambda=-100$ where the procedure of obtaining solutions is very difficult technically, and we could find them only to
$\text{BE}\approx 0.21$; this corresponds to the leftmost point in the graph).

It is seen from Fig.~\ref{fig_mass_E} that for all  $\bar \lambda$ there is a maximum of the mass at some value of
$\bar E$ (or  $\bar g_c$). Such a behavior of the curves resembles the behavior of the corresponding
``mass~--~central density'' dependencies for boson stars supported by a complex scalar field
(see, e.g., Refs.~\cite{Colpi:1986ye,Gleiser:1988ih,Herdeiro:2017fhv}). In the case of boson stars, the presence of such a maximum corresponds to the boundary between configurations
which are stable or unstable against linear perturbations~\cite{Gleiser:1988ih}. Naively, one might expect that for the Dirac stars a similar situation will occur.
But this issue requires special studies.

\subsection{Limiting configurations for $|\bar \lambda| \gg 1$}

It was shown in Ref.~\cite{Herdeiro:2017fhv} that the maximum mass of Dirac stars supported by a noninteracting spinor field
 is $M^{\text{max}}\approx 0.709 M_p^2/\mu$.
For the mass of a spinor field $\mu\sim 1~\text{GeV}$, it gives the total mass $M\sim 10^{14}~\text{g}$, i.e., the stars with small masses and radii
$R\sim 10~\text{fm}$.
(Then, by analogy to miniboson stars \cite{Lee:1988av}, one can speak of mini-Dirac stars.) In turn, one can see from the results obtained above that the use of positive values
of the coupling constant $\bar \lambda$ leads to decreasing the maximum mass. In this connection, from the point of view of possible astrophysical applications,
it seems more interesting to use negative values of $\bar \lambda$. Numerical calculations indicate that as $|\bar \lambda|$ increases
there is a considerable growth in maximum masses of the configurations under consideration
(see Fig.~\ref{fig_mass_E}). For clarity,
in Fig.~\ref{fig_mass_lambda}, we have plotted the dependence of the maximum mass
 $M^{\text{max}}$ as a function of $|\bar \lambda|$. In this figure, the solid line corresponds to the interpolation formula
 \begin{equation}
\label{M_max_approx}
	M^{\text{max}}\approx 0.415 \sqrt{|\bar \lambda|}M_p^2/\mu,
\end{equation}
which holds asymptotically for $|\bar \lambda|\gg 1$.

\begin{figure}[t]
	\begin{minipage}[t]{.49\linewidth}
		\begin{center}
			\includegraphics[width=1\linewidth]{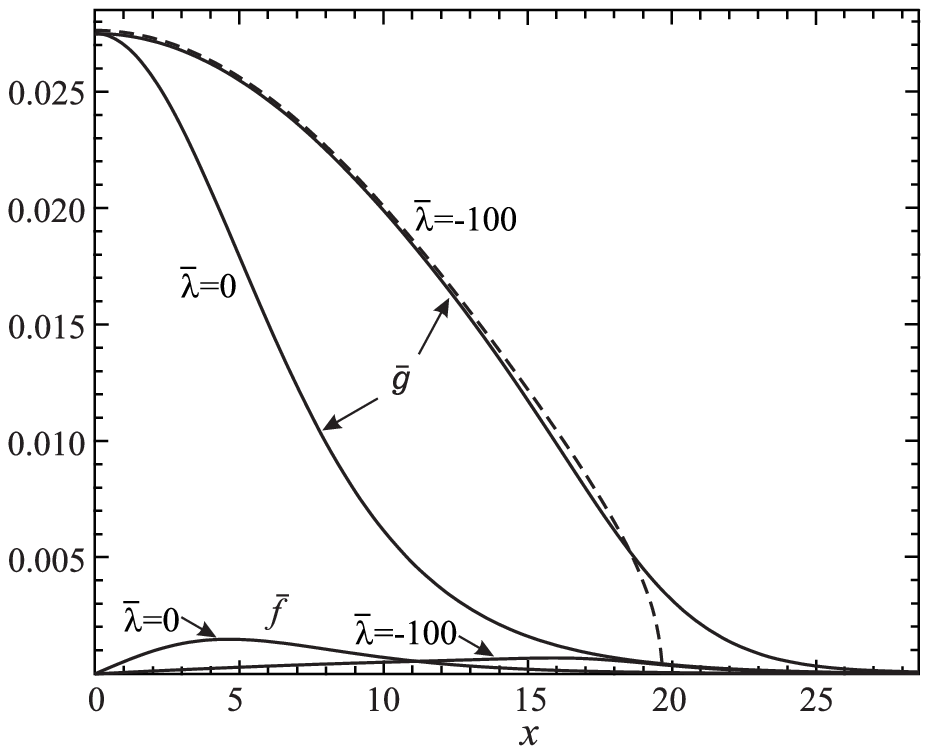}
		\end{center}
\vspace{-0.5cm}
		\caption{
		Spinor fields $\bar g$ and $\bar f$  as functions of dimensionless radius $x$ for $\bar \lambda=-100$ and $\bar \lambda=0$.
The dashed line shows the solution to Eqs.~\eqref{g_approx}-\eqref{fieldeqs-4_dmls_approx}  with $\bar E/\sigma_c$
from the exact $\bar g_c=0.0275, \bar \lambda=-100$ model, scaled to $\bar \lambda=-100$.
		}
		\label{fig_field_distr}
	\end{minipage}\hfill
\begin{minipage}[t]{.49\linewidth}
		\begin{center}
			\includegraphics[width=1\linewidth]{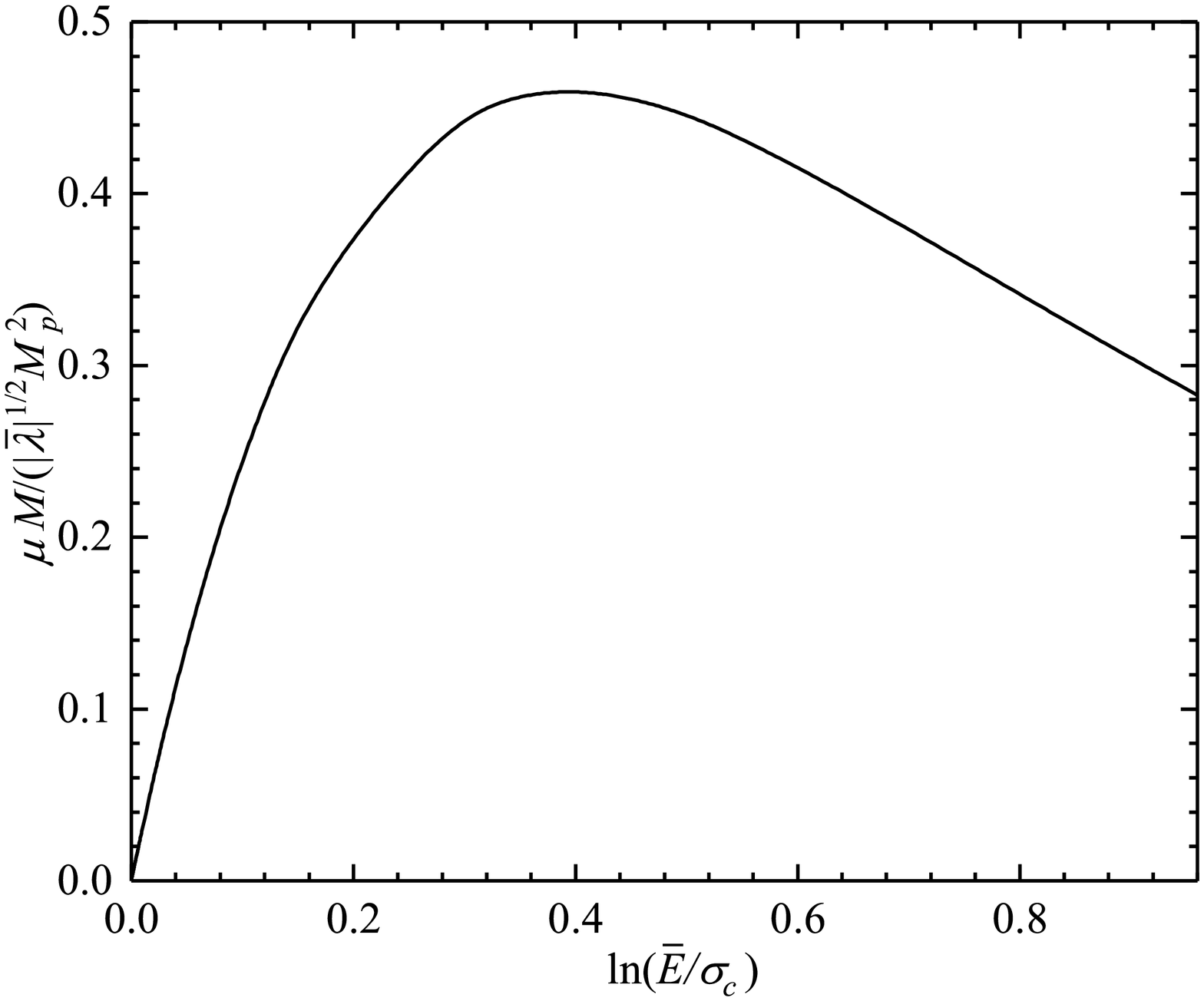}
		\end{center}
\vspace{-0.5cm}
		\caption{
		Dimensionless Dirac-star total mass $\bar M_*$ as a function of $\bar E/\sigma_c$ for the limiting configurations described by Eqs.~\eqref{g_approx}-\eqref{fieldeqs-4_dmls_approx}.
The graph is plotted only for the values of $\bar E$ for which the binding energy is positive.
 		}
		\label{fig_mass_E_sigma}
	\end{minipage}
\end{figure}

We have found that in the case of the spinor systems considered here,
as in the case of boson stars of Ref.~\cite{Colpi:1986ye}, the large-$|\bar \lambda|$ configurations have the structure that differs significantly from that of the
small-$|\bar \lambda|$ systems. These distinctions are illustrated in Fig.~\ref{fig_field_distr}, which shows
the spinor field distributions along the radius for the cases of $\bar \lambda=0$ and $\bar \lambda=-100$
(for the case of $\bar \lambda=-100$, we take the configuration with a maximum mass marked by a bold dot in Fig.~\ref{fig_mass_E}).
It is seen from this figure that in both cases the main contribution to the energy density (and correspondingly to the mass) is given by the function
 $\bar g$. This function tends exponentially to zero in a characteristic length of $1/\mu$ for small
$|\bar \lambda|$ but for large $|\bar \lambda|$ is characterized by relatively slow decline out to radii $\sim 2 |\bar \lambda|^{1/2}/\mu$ with exponential decay only at larger radii (cf. Ref.~~\cite{Colpi:1986ye}).
For this reason, the majority of the mass of the large-$|\bar \lambda|$ systems is concentrated in the region
of slow decline, which becomes increasingly dominant as $|\bar \lambda|$ increases.

As in the case of boson stars of Ref.~\cite{Colpi:1986ye}, such a behavior of the spinor fields enables one to introduce an alternative nondimensionalization
of Eqs.~\eqref{fieldeqs-1_dmls}-\eqref{fieldeqs-4_dmls} valid at large
$|\bar \lambda|$: $\bar g_*,
\bar f_*=|\bar \lambda|^{1/2}\bar g,
\bar f, \bar m_*=|\bar \lambda|^{-1/2}\bar m
$, and $x_*=|\bar \lambda|^{-1/2}x$.
Using these new variables and taking into account that the leading term in Eq.~\eqref{fieldeqs-1_dmls} is the third term
$(\ldots)\bar g$, this equation yields (in the approximation of $\bar f \ll \bar g$)
\begin{equation}
 \label{g_approx}
 \bar g_* = \sqrt{-\frac{1}{8}
 \left(1 - \frac{\bar E}{\sigma\sqrt{N}}\right)}.
\end{equation}
Substituting this expression into Eqs.~\eqref{fieldeqs-3_dmls} and \eqref{fieldeqs-4_dmls}, we have (to the same accuracy)
\begin{eqnarray}
	&&\frac{d \bar m_*}{d x_*} = 8 x_*^2 \bar g_*^2\left(
		\frac{\bar E}{\sigma\sqrt{N}} - 4\bar g_*^2
	\right),
\label{fieldeqs-3_dmls_approx}\\
	&&\frac{d \sigma}{d x_*} = 8\bar E x_*\frac{\bar g_*^2}{N^{3/2}},
\label{fieldeqs-4_dmls_approx}
\end{eqnarray}
where now $N=1-2\bar m_*/x_*$. As  $|\bar \lambda|$ increases, the accuracy of
Eqs.~\eqref{g_approx}-\eqref{fieldeqs-4_dmls_approx} becomes better.
In particular, when $\bar \lambda=-100$, from comparison  of the exact and approximate solutions, one can observe their good agreement,
except the behavior at large radii (see Fig.~\ref{fig_field_distr}). Since
$\bar \lambda$ does not appear explicitly in Eqs.~\eqref{fieldeqs-3_dmls_approx} and \eqref{fieldeqs-4_dmls_approx}, one can use these limiting equations to determine the rescaled total mass
$\bar M_*= M/\left(|\bar \lambda|^{1/2}M_p^2/\mu\right)$ as a function of the single free parameter $\bar E/\sigma_c$.
The corresponding results of numerical solution of Eqs.~\eqref{g_approx}-\eqref{fieldeqs-4_dmls_approx} are given in Fig.~\ref{fig_mass_E_sigma},
from which one can see the presence of a maximum of the mass,
\begin{equation}
\label{M_max_approx_2}
	M_*^{\text{max}} \approx 0.41 \sqrt{|\bar \lambda|} M_p^2/\mu.
\end{equation}
One can see that this expression agrees very well with that given in Eq.~\eqref{M_max_approx}, and this confirms that the above approximation is in good
agreement with the exact solution.

As in the case of boson stars of Ref.~\cite{Colpi:1986ye},
 for the Dirac star, the ground state of the spinor field is {\it not} a zero-energy state (because of self-gravity). Moreover, at large $|\bar \lambda|$, the spinor field is spread over a relatively large length scale
$|\bar \lambda|^{1/2}\mu^{-1}\gg \mu^{-1}$; this enables one to neglect locally the derivatives of
 $\bar g$ and $\bar f$. This allows the possibility of, first, obtaining the solution of the equation for the spinor field
  \eqref{fieldeqs-1_dmls} in the form of \eqref{g_approx} when one can neglect the influence of the function
  $\bar f$ and its derivative. Second, in the approximation of neglecting the derivatives, one can introduce an effective equation of state.
  To do this, let us use the components $T_0^0=\varepsilon$ and $T_1^1=-p_r$ of the energy-momentum tensor \eqref{EM}, where $\varepsilon$ is the effective energy density of the spinor fluid and
$p_r$ is its radial pressure,
\begin{eqnarray}
	&&\bar \varepsilon\equiv \frac{\varepsilon}{\gamma} = \frac{8\bar E}{\sigma\sqrt{N}} \left(\bar f^2 + \bar g^2\right) +
	32\bar \lambda\left(\bar f^2-\bar g^2\right)^2,
\label{T00}\\
	&&\bar p_r\equiv\frac{p_r}{\gamma} = 8\sqrt{N}
	\left(\bar g\bar f^\prime - \bar f\bar g^\prime\right) -
	32\bar \lambda\left(\bar f^2 - \bar g^2\right)^2
\label{T11}
\end{eqnarray}
with $\gamma=c^2 M_p^2/\left(4\pi \mu\lambda_c^3\right)$.
(Note that, as in the case of boson stars of Ref.~\cite{Colpi:1986ye}, the radial, $p_r$, and tangential,
 $p_t=-T_2^2$, components of pressure for the Dirac star are not equal to each other.) For $|\bar \lambda|\gg 1$, in the approximation used here, one can obtain
$$
\bar\varepsilon_*\equiv |\bar \lambda|\bar\varepsilon=8\bar g_*^2\left(\frac{\bar E}{\sigma\sqrt{N}}+4\bar g_*^2\right),\quad
\bar p_{r*}\equiv |\bar \lambda|\bar p_r=32 \bar g_*^4.
$$
Taking into account the expression  \eqref{g_approx} for $\bar g_*$ and eliminating from these relations
 $\bar E/\left(\sigma\sqrt{N}\right)$, one can derive the following effective equation of state:
\begin{equation}
\label{EoS_eff}
	\bar p_{r*} = \frac{1}{9} \left(
		1 + 3\bar\varepsilon_*\pm\sqrt{1 + 6\bar\varepsilon_*}
	\right).
\end{equation}
The dimensionless quantities appearing here are related to the dimensional energy density and pressure in the following manner:
$\bar p_{r*}, \bar\varepsilon_*= (p_r, \varepsilon)/\varepsilon_0$, where $\varepsilon_0 = \left(\mu c^2\right)^2/|\lambda|$.
Then, the relations \eqref{M_max_approx} and \eqref{M_max_approx_2}, using the expression for
 $\bar \lambda$ from Eq.~\eqref{dmls_var}, are equivalent to the statement that $M^{\text{max}}\sim M_p^3/\sqrt{\varepsilon_0}$
 for a fluid star with an equation of state of the form of Eq.~\eqref{EoS_eff} (cf. Ref.~\cite{Colpi:1986ye} where a similar expression has been
 obtained for boson stars). In the case of boson stars, such a limiting transition from a
 scalar field configuration to a fluid system when the coupling constant tends to infinity
 enables one to assume that stable configurations can occur. In fact, both the systems supported by
 a relativistic fluid~\cite{Tooper2} and the configurations consisting of a complex scalar field~\cite{Gleiser:1988ih},
 located to the left of the first peak in the mass in the ``mass--central density'' diagram,
 are stable against linear perturbations. It seems reasonable to suppose
that the same stability criterion may be applied for the spinor field configurations
considered in the present paper. However, this question requires special studies, for example, by analogy with Ref.~\cite{Finster:1998ws}.

We conclude this section with the expression for the effective pressure \eqref{T11}, which, by changing the derivatives
$\bar f^\prime$ and $\bar g^\prime$ using Eqs.~\eqref{fieldeqs-1_dmls} and \eqref{fieldeqs-2_dmls}, respectively, can be rewritten as
\begin{equation}
\label{p_r_eff}
\bar p_r=8\left[
\frac{\bar E}{\sigma\sqrt{N}} \left(\bar f^2 + \bar g^2\right)+\left(\bar f^2 - \bar g^2\right)-2\frac{\bar f\bar g}{x}+4\bar \lambda\left(\bar f^2 - \bar g^2\right)^2
\right].
\end{equation}
This expression permits us to see the physical meaning of the coupling constant $\bar\lambda$: the case of $\bar\lambda > 0$ corresponds to the attraction,
and the case of $\bar\lambda < 0$ corresponds to the repulsion. Correspondingly, in the case of negative $\bar\lambda$, the self-interaction term ensures a counterbalance
force to the gravitational attraction; this eventually enables us to get configurations with large masses. In turn, the numerical computations indicate that the magnitudes of the effective pressure
\eqref{p_r_eff} and the pressure gradient $d \bar p_r/dx$ along the radius of the configuration are determined by specific values of the system parameters $\{\bar\lambda, \bar g_c\}$:
for each of these pairs, the pressure $\bar p_r$ can be positive (negative) with negative (positive) gradient, or the pressure and the gradient are alternating functions along the radius.

\section{Conclusions and discussion}
\label{concl}

The paper studies compact strongly gravitating configurations supported by nonlinear spin-$\frac{1}{2}$ fields.
The use of two spinor fields having opposite spins enabled us to get a diagonal energy-momentum tensor suitable for a description of spherically symmetric systems.
Consistent with this, we have found localized regular zero-node asymptotically flat solutions for explicitly time-dependent spinor fields,
oscillating with a frequency $E/\hbar$. It was shown that for all values of $E$ and of the coupling constant $\lambda$ considered here
these solutions describe configurations possessing a positive ADM mass. This enables one to use such solutions for a description of compact gravitating objects (Dirac stars).

From the results obtained earlier for Dirac stars {\it without} nonlinearity,
it follows that for typical values of the spinor field mass total masses of such configurations are extremely small
(see, e.g., Ref.~\cite{Herdeiro:2017fhv}). Here, we show that the
 presence of nonlinearity of spinor fields can alter the situation drastically.
 In the simplest case, the nonlinearity can be chosen in a quadratic form of the type $\lambda (\bar \psi \psi)^2$.
 Then, families of gravitational equilibria may be parametrized by the single dimensionless quantity
$\bar \lambda =\lambda M_p^2 c/4\pi \hbar^3$. Consistent with the dimensions of
 $[\lambda]=\text{erg cm}^3$, one can assume that its characteristic value is
  $\lambda \sim  \tilde \lambda \,\mu c^2 \lambda_c^3$, where the dimensionless quantity $\tilde \lambda \sim 1$.
  Then, the dependence of the maximum mass of the systems under consideration on
  $|\bar \lambda|$  in the limit  $|\bar \lambda|\gg 1$  [see Eq.~\eqref{M_max_approx}] can be represented as
$$
	M^{\text{max}}\approx 0.415 \sqrt{|\bar \lambda|}\frac{M_p^2}{\mu} \approx 0.19\, \sqrt{|\tilde \lambda|}M_{\odot}\left(\frac{\text{GeV}}{\mu}\right)^2.
$$
This mass is comparable to the Chandrasekhar mass for the typical mass of a fermion $\mu\sim 1~\text{GeV}$.
In this respect, the behavior of the dependence of the maximum mass of the Dirac stars on the coupling constant is similar to that of boson stars of Refs.~\cite{Colpi:1986ye, Mielke:2000mh}.

Note that, in the absence of gravity, a nonlinear spinor field has been investigated in Ref.~\cite{heis} relating to the problem of the quantization of an electron.
In Refs.~\cite{Finkelstein:1951zz,Finkelstein:1956}, it was shown that the corresponding nonlinear Dirac equation has regular solutions
with finite energy (also without gravity).  This permits us to assume that in our case a nonlinear spinor field can approximately describe fermions (or quarks)
which are in some quantum state where they can be approximately described by some collective wave function obeying a nonlinear Dirac equation.
A similar situation occurs in considering bosons in a Bose-Einstein condensate described by the  Gross-Pitaevski equation
and also in describing Cooper pairs in a superconductor by means of the Ginzburg-Landau equation.

In conclusion, we would like to briefly address the question of stability of the configurations under consideration.
Similarly to models of neutron and boson stars, which can be parametrized by their central
densities, one can consider a one-parameter family of Dirac stars described by the central value of the spinor field $g_c$. The total mass is then a function of this parameter,
and for any value of the coupling constant $\lambda$, there exists a first peak in the mass (a local maximum).
In the case of neutron and boson stars, a transition through this local maximum indicates an onset of instability against perturbations which compress the entire
star as a whole.
One can naively expect that in the case of Dirac stars a similar situation will also take place. However, this requires special consideration by investigating the stability of
spinor field configurations against, for instance, linear perturbations.

\section*{Acknowledgments}
The authors gratefully acknowledge support provided by Grant No.~BR05236494
in Fundamental Research in Natural Sciences by the Ministry of Education and Science of the Republic of Kazakhstan. We are grateful to the Research Group
Linkage Programme of the Alexander von Humboldt Foundation for the support of this research and also would like to thank the Carl von Ossietzky University of
Oldenburg for hospitality while this work was carried out. We also wish to thank E.~Radu for a fruitful discussion of the problem statement and of the obtained results.


\begin{thebibliography}{99}
\bibitem{AmenTsu2010}
 L. Amendola and S. Tsujikawa, {\it  Dark Energy: Theory and Observations} (Cambridge University Press,
Cambridge, England, 2010).

\bibitem{Linde:2007fr}
  A.~D.~Linde,
  Lect.\ Notes Phys.\  {\bf 738}, 1 (2008).

\bibitem{Schunck:2003kk}
  F.~E.~Schunck and E.~W.~Mielke,
  Classical Quantum Gravity  {\bf 20}, R301 (2003).

\bibitem{Bartnik:1988am}
  R.~Bartnik and J.~Mckinnon,
  Phys.\ Rev.\ Lett.\  {\bf 61}, 141 (1988).

\bibitem{Brito:2015pxa}
  R.~Brito, V.~Cardoso, C.~A.~R.~Herdeiro, and E.~Radu,
  Phys.\ Lett.\ B {\bf 752}, 291 (2016).

\bibitem{Finster:1998ws}
  F.~Finster, J.~Smoller, and S.~T.~Yau,
  Phys.\ Rev.\ D {\bf 59}, 104020 (1999).

\bibitem{Herdeiro:2017fhv}
  C.~A.~R.~Herdeiro, A.~M.~Pombo, and E.~Radu,
  Phys.\ Lett.\ B {\bf 773}, 654 (2017).

\bibitem{Bronnikov:2004uu}
  K.~A.~Bronnikov, E.~N.~Chudaeva, and G.~N.~Shikin,
  Gen.\ Relativ.\ Gravit.\  {\bf 36}, 1537 (2004).

\bibitem{Bronnikov:2009na}
  K.~A.~Bronnikov and J.~P.~S.~Lemos,
  Phys.\ Rev.\ D {\bf 79}, 104019 (2009).

\bibitem{Saha:2014hea}
B.~Saha,
Astrophys.\ Space Sci.\  {\bf 357},  28 (2015).

\bibitem{Saha:2015sna}
B.~Saha,
Eur.\ Phys.\ J.\ Plus {\bf 130},  208 (2015).

\bibitem{Saha:2015rna}
B.~Saha,
Eur.\ Phys.\ J.\ Plus {\bf 131}, 170 (2016).

\bibitem{Saha:2016ebv}
B.~Saha,
Int.\ J.\ Theor.\ Phys.\  {\bf 55}, 2259 (2016).

\bibitem{Saha:2016cbu}
B.~Saha,
Eur.\ Phys.\ J.\ Plus {\bf 131},  242 (2016).


\bibitem{Krechet:2014nda}
  V.~G.~Krechet and I.~V.~Sinilshchikova,
  Russ.\ Phys.\ J.\  {\bf 57},  870 (2014).

 \bibitem{Adanhounme:2012cm}
  V.~Adanhounme, A.~Adomou, F.~P.~Codo, and M.~N.~Hounkonnou,
  J.\ Mod.\ Phys.\  {\bf 3}, 935 (2012).

\bibitem{Ribas:2010zj}
M.~O.~Ribas, F.~P.~Devecchi, and G.~M.~Kremer,
Europhys.\ Lett.\  {\bf 93}, 19002 (2011).

\bibitem{Ribas:2016ulz}
M.~O.~Ribas, F.~P.~Devecchi, and G.~M.~Kremer,
  Mod.\ Phys.\ Lett.\ A {\bf 31},  1650039 (2016).

\bibitem{Dzhunushaliev:2018bdi}
  V.~Dzhunushaliev,
  Gravitation Cosmol.\  {\bf 24},  267 (2018).

\bibitem{ArmendarizPicon:2003qk}
  C.~Armendariz-Picon and P.~B.~Greene,
  Gen.\ Relativ.\ Gravit.\  {\bf 35}, 1637 (2003).

\bibitem{Lawrie2002}
I.~Lawrie, {\it A Unified Grand Tour of Theoretical Physics} (Institute of Physics Publishing, Bristol, 2002).

\bibitem{Li:1982gf}
X.~z.~Li, K.~l.~Wang, and J.~z.~Zhang,
  Nuovo Cimento\ A {\bf 75}, 87 (1983).

\bibitem{Li:1985gf}
K.~L.~Wang and J.~Z.~Zhang,
  Nuovo Cimento\ A {\bf 86}, 32 (1985).

\bibitem{obukhov}
Yu. Obukhov,
(to be published).

\bibitem{Colpi:1986ye}
  M.~Colpi, S.~L.~Shapiro, and I.~Wasserman,
  Phys.\ Rev.\ Lett.\  {\bf 57}, 2485 (1986).

\bibitem{Gleiser:1988ih}
  M.~Gleiser and R.~Watkins,
  Nucl.\ Phys.\  {\bf B319}, 733 (1989).

\bibitem{Lee:1988av}
  T.~D.~Lee and Y.~Pang,
  Nucl.\ Phys.\  {\bf B315}, 477 (1989).

\bibitem{Tooper2}
  R.~Tooper,
  Astrophys.\ J.\  {\bf 142}, 1541 (1965).

\bibitem{Mielke:2000mh}
  E.~W.~Mielke and F.~E.~Schunck,
  Nucl.\ Phys.\  {\bf B564}, 185 (2000).

\bibitem{heis}
	W.~Heisenberg, {\it Introduction to the Unified Field Theory of Elementary Particles.} (Interscience Publishers, London, 1966).

\bibitem{Finkelstein:1951zz}
  R.~Finkelstein, R.~LeLevier, and M.~Ruderman,
  Phys.\ Rev.\  {\bf 83}, 326 (1951).

\bibitem{Finkelstein:1956}
R. Finkelstein, C. Fronsdal, and P. Kaus,
  Phys.\ Rev.\  {\bf 103}, 1571 (1956).

\end{thebibliography}
\end{document}